\newcount\pageformat\pageformat=2  
\newcount\pssize

\ifnum\pageformat=2
 \documentclass[aps,prb,twocolumn,floatfix,showpacs,amsmath,amssymb]{revtex4}
\else
 \documentclass[aps,prb,preprint,floatfix,showpacs,amsmath,amssymb]{revtex4}
\fi
\usepackage{graphicx}

\begin{document}
\title{Signs of Phase Transitions in Two-Nucleon Systems}

\author{B. F. Kostenko} \email{bkostenko@jinr.ru}

\affiliation{Joint Institute for Nuclear Research, Dubna\\
141980 Moscow region, Russia}

\author{J. Pribi\v{s}}

\affiliation{Technical University, Ko\v{s}ice,  Slovak Republic}

\date{\today }

\begin{abstract}
The properties of dense nuclear matter under extreme conditions are
a subject of the large experimental activities worldwide.  Heavy-ion
collision experiments are the agenda item at RHIC, LHC, FAIR, and
NICA facilities. Meanwhile, a complementary approach to the
heavy-ion collision researches devoted to investigation of phase
transitions in few-nucleon systems has not been discussed. In this
paper, we try to fill up the gap. It is shown that signals of the
phase transition of deuteron into 6-q bag as well as signs of
formation of the pion Bose-Einstein condensate in compressed
two-nucleon systems might be already observed in  deuteron-deuteron
collisions.
\end{abstract}


\maketitle

\section{\label{sec1}Introduction}
High-energy nuclear collisions allow the study of new phases of
nuclear matter under extreme conditions at which the phase
transition of nuclear matter  to a color-deconfined state was
predicted by the fundamental theory of strong interactions, the
Quantum Chromodynamics (QCD). The experimental programs  at BNL and
CERN have already confirmed that the extreme conditions of matter
necessary to reach the new phase can be reached in the high-energy
nuclear collisions. However, identifying and studying the properties
of those phases is a challenging task, mainly because of many-body
effects and nonperturbative nature of the processes involved. These
challenges stimulate putting forward new experimental and
theoretical ideas aimed at search of unambiguous signatures of the
phase transition onset. Recently a proposal of QCD investigation at
high density and low temperature complementary to the high-energy
heavy nuclear collisions was suggested \cite{Kostenko1, Kostenko2}.
The proposal is based on the fact that a large number of nucleons in
the interaction region is not  necessary for the phase transition to
occur, and only a change of the vacuum state should be initiated by
some experimental environment. Detection of two- and three-nucleon
short range correlations \cite{CLAS} affords an opportunity to use
the dense few-nucleon correlated systems  of this type (SRC) as
targets which correspond to small fragments of nuclear matter in the
dynamically broken chiral symmetry states. Collisions of SRC with
bombarding particles can initiate the chiral phase transition,
ending in the creation of a multi-baryon(MB). Thus, the observation
of MB would be a direct evidence of the chiral condensate
disappearance and the chiral symmetry restoration in the interaction
area. Separation of a MB mass from the secondary particle background
is feasible if the MB decay width is narrow enough.  That requires
the excitation energy of produced MB to be low. For this purpose, it
is reasonable to select only those experimental events in which the
MB creation is accompanied with a high momentum particle, taking
away an essential part of the energy from the interaction region (a
cooling effect). In this paper, we focus on new developments in this
direction outlined in \cite{Kostenko1, Kostenko2} and put them in a
context with some of older experimental data  taken at JINR
synchrophasotron \cite{Baldin, Troyan1, Troyan2, Troyan3}.

\begin{figure}
\begin{center}
\includegraphics[width=8.6cm]{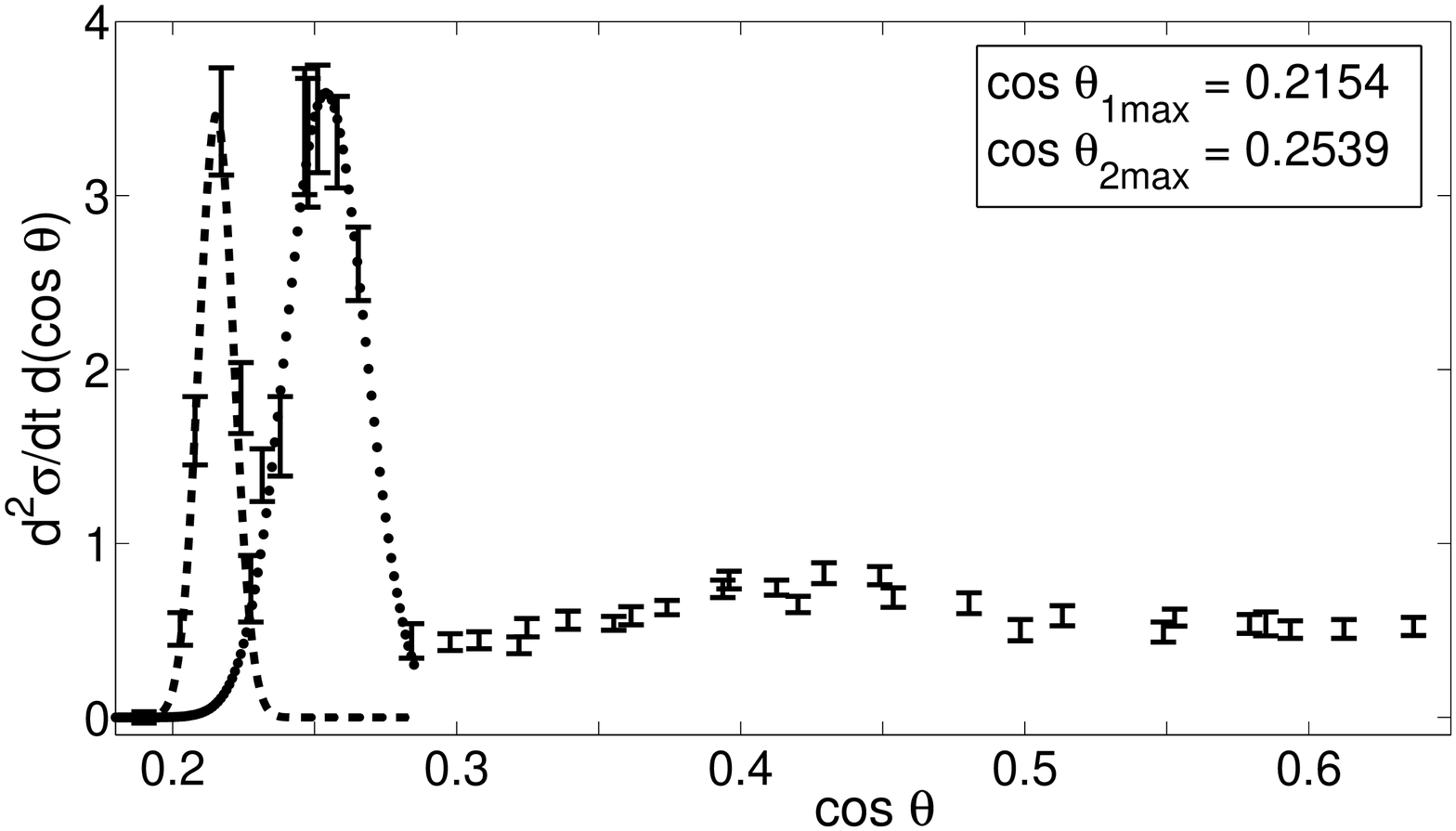}
\end{center}
\caption{\small Double differential cross-sections, in
mb$\cdot$c$^2$/GeV$^2$, of the D$+$D$\to$M$_X+$D reaction against
cosine of the target nucleus scattering angle \cite{Baldin}.  The
dashed and dotted curves correspond to approximations of the first
two peaks by the Gaussian functions which maxima positions are given
in the insertion.}\label{Exp}
\end{figure}

An experiment \cite{Baldin} was designed for measurement of the
cross-sections of elastic pp-, ND-, and DD-scattering  at 8.9 GeV
momentum of primary protons and deuterons. Particularly, three peaks
were observed in the spectrum of the missing masses of the reaction
D$+$D$\to$M$_X+$D at $t=-0.495$ GeV$^2$ (see Fig.~\ref{Exp}).
Till now the first of them corresponding to the most heavy M$_X$ was
estimated to cover the elastic DD scattering; the second one was
interpreted as a manifestation of the scattering of a projectile
deuteron's nucleon by the target deuteron; in regard to the third
peak, it was suggested to appear because of 1) a contribution of the
constituent quark scattering, 2) a contribution of an excited state
of deuteron (e.g.,  6q-bag), and 3) to be a kinematic manifestation
of a baryon N$^*$ with a value of mass in the neighborhood of 1400
MeV.

Experimental findings occurred after the paper \cite{Baldin} was
written give cause for re-examination of its conclusions. Data from
\cite{Troyan1, Troyan2} employing 38915 events  will play an
especially important role in our consideration. So far as the
interpretations of the third peak promise detection of the chiral
phase transition,  we begin  with it. Thereafter problems concerning
the first two peaks will be discussed.

\section{\label{sec2} Possible dibaryons in the third peak region}
\begin{figure}
\begin{center}
\includegraphics[width=8.6cm]{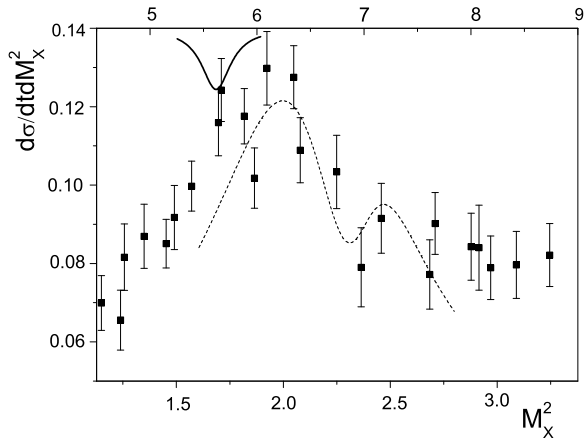}
\end{center}
\caption{\small The experimental data (dots) in the range of the
third peak and their explanation by the sum of contributions of
N$+$D$\to$X$+$D reactions (dashed line).  The top scale corresponds
to the kinematics of reaction D$+$D$\to$X$+$D  which implies the
dibaryon production.  A possible contribution of a dibaryon at 2.37
GeV, $\Gamma\approx$ 70 MeV, reported by WASA-at-COSY Collaboration
\cite{WASA} into the third region is  shown with the overturned
solid line.}\label{Peak3}
\end{figure}
Kinematics of the N$+$D$\to$X$+$D reaction reads\footnote{All
kinematic kinematic relations given in this paper can be found as
follows. Let us denote by 1+2 $\to$ 3+4 a reaction at issue, where
the projectile, target and registered particles are designated by 1,
2 and 4, correspondingly, and 3 denotes an object X which mass
should be determined. Two different expressions for the Lorentz
invariant Mandelstam variable $u$, $u=(p_1 - p_4)^2=(p_2 - p_3)^2$,
$p_i=(E_i,\mathbf{P}_i)$, $\left| \mathbf{P}_i \right| = \sqrt
{E_i^2 - M_i^2 }$, allow to connect $M_X$ and $\cos \theta$. The
energy of particle 4 as function of $M_2,\; M_4$ and $t$ may be
found by making use of a relation $t=(p_2 - p_4)^2$. In addition,
$E_3=E_1+E_2-E_4$.}
\begin{equation}\label{form1}
{{\it M_X}}^{2}={{\it M_N}}^{2}+t+
\end{equation}
$$
+1/2\,{\frac {\sqrt {{{\it P_1}}^{ 2}+4\,{{\it M_N}}^{2}}t+{\it
P_1}\,\sqrt {t \left( -4\,{{\it M_D}}^ {2}+t \right) }\cos \theta
}{{\it M_D}}},
$$
where $P_1$ is momentum of the primary deuteron, $P_1=8.9$, and
$M_N=0.94$, $M_D=1.8756$ GeV. Here X may be any baryon resonance or
nucleon escorted by one or more pions. In framework of this model,
the differential cross-section $d^2 \sigma/dt dM_X^2$ in a region
$M_X \approx 1.4$ GeV may be expressed as follows:
\begin{equation}\label{cross}
C_0+\sum\limits_{i =
1}^3 {C_i f_i }  + 2\sum\limits_{\begin{array}{*{20}c}
   {i,j = 1}  \\
   {i < j}  \\
 \end{array} }^3 {k_{ij} \sqrt {C_i C_j f_i f_i } \cos (\varphi_i  - } \varphi_j
 ),
\end{equation}
where resonances N(1440), N(1520), and N(1535) were taken into
account\footnote{Production of $\Delta (1232)$ is forbidden by the
isospin conservation law.}. We have chosen functions $f_i$  in the
form\cite{Pilk}  $f_i= M_i \Gamma_i/((M_X^2-M_i^2)^2+\Gamma_i^2
M_i^2)$. Data on baryon resonances were taken from ref. \cite{PDG};
a contribution of the reaction with nucleon and several pions in the
final state was approximated by a constant; values of parameters in
(\ref{cross}) were found to obtain the best description of the data,
according to a global optimization procedure. We have introduced for
the interference terms factors $k_{ij} \in [0,1]$ which take into
account a value of indistinguishability of two different
resonances\footnote{Constraints imposed by the angular momentum and
parity conservation laws give overlapping but nonidentical regions
for different baryon resonances.}. In Fig.~\ref{Peak3}, an attempt
to explain a fine structure in the range of the third peak by the
sum of contributions of reactions N$+$D$\to$X$+$D\footnote{Values of
parameters in (\ref{cross}) were found as follows: $C_0=$0.00419,
$C_1=$ 0.05239, $C_2=$0.02371, $C_3=$0.02560, $k_{12}=$0.84952,
$k_{13}=$ 0.78284, $k_{23}=$1, $\varphi_1=$0, $\varphi_2=$3.17519,
$\varphi_3$ = 0.04110.}. One can see that, in principle, it possible
to explain at least a part of the fine structure in the third peak
region by the sum of contributions of the processes with nucleon
excitations.

For the dibaryon production in the reaction D$+$D$\to$2B$+$D, the
isospin conservation  leads to $I_{2B}=0$. The kinematics,
\begin{equation}\label{form2}
M_X^2  = M_D^2  + t + \frac{{E_1 t + P_1 \sqrt {t( - 4M_D^2  + t)}
\cos \theta }}{{M_D }},
\end{equation}
states that the fine structure in the third peak region is described
if one supposes the existence of a dibaryon at $M_{2B} \approx  $
2.38  GeV (see the upper scale in Fig.~\ref{Peak3}). It looks like
it was recently reported\cite{WASA} by WASA-at-COSY Collaboration in
reaction pp $\to$ D $\pi^0 \pi^0$ at 2.37 GeV, $\Gamma\approx$ 70
MeV. Similar masses were found in a pp$\pi^+$ system in
\cite{Troyan3}. Therefore, it is plausible to expect that these
hypothetic dibaryons decay into two nucleons and one or two pions.

It is interesting to check if the fine structure in the third peak
region  may be explained as constituent quark scattering in reaction
qD$\to$qD. The elastic scattering of a constituent quark by the
target deuteron may be considered in the framework of a model in
which values of momentum and mass of the projectile quark are
considered in the form
$$
P_q = x P_1, \qquad M_q = x M_D ,
$$
where  $x$ is determined from kinematics of the reaction. The model
gives for $ \cos \theta = 0.396$  corresponding to $M_{2B} \approx $
2.38 GeV, a value of quark mass
\[
M_q  = \frac{{ - M_D^2 t}}{{E_1 t + P_1 \sqrt {t( - 4M_D^2  + t)}
\cos \theta }},
\]
about 0.351  GeV. It contradicts to modern constituent quark models;
see, e.g., ref. \cite{Jovan} in which $M_q = 0.318$ GeV.

\section{\label{sec3}The first two peaks' puzzle }
The Gaussian two-peak approximation results in $\cos \theta_1 =
0.2154$ and $\cos \theta_2 = 0.2539$ for the location of the first
two peaks' maxima (see Fig.~\ref{Exp})\footnote{It is shown below
that the Gauss distribution might arise from a sum of many near
resonances. An extension of statistics  may modify slightly the
overall distribution.}. It was very unexpected to find that elastic
D-D scattering gives the angle distribution with a maximum at
0.2272, see (\ref{form2}) for $M_X=M_D$, i.e. between $\cos
\theta_1$ and $\cos \theta_2$. Similarly, elastic N-D scattering
described by (\ref{form1}) with $M_X=M_N$ has a maximum at 0.2661,
clearly shifted from the second peak location. Thus, the explanation
of the first two peaks by means of contributions of the elastic D-D
and N-D scattering fails and their origin remains unclear. At first
glance, the discrepancy may be attributed to systematic errors
committed in the experiment, but a subsequent calculations found out
that another astonishing explanation is more plausible.

To explain positions of the first two peaks, different models have
been tried out. The models were based on the fact that only  the
recoil deuteron was unambiguously identified in\cite{Baldin} but
masses of all other participants were unknown. Therefore, any
transitions X$+$Y$\to$Z$+$D are allowed to be taken into account.
For example, a scattering X$+$D $\to$ D$+$D explains the first peak
location if one  assigns to X a value of mass of about 1913 Mev
which turns out to be close to 1916$\pm$2 MeV, observed in a pp
dibaryon spectrum by Yu.A. Troyan \cite{Troyan1, Troyan2}. A model
D$+$D $\to$ X$+$D gives for the second peak location if one assumes
$M_X =1965$ MeV. The data from \cite{Troyan1, Troyan2} contain a
corresponding dibaryon with $M_X =1965 \; \pm 2$ MeV.

Analysis of other models  showed that almost each dibaryon observed
in \cite{Troyan1, Troyan2} can give a contribution to the first two
peaks observed in \cite{Baldin}, under an assumption that masses of
dibaryons detected in the np-system are 1 MeV less than the
corresponding masses in the pp-system. In Table~\ref{tab1},
considered reactions are shown in the first column. The second
column specifies masses of ingoing or outgoing objects in the
deuteron scattering experiment \cite{Baldin}. Dibaryon masses found
for the pp-system in refs. \cite{Troyan1, Troyan2} are given in the
third column.
\begin{table}\small
\caption{\label{tab1}\small Kinematically admissible masses (KAM)
which might contribute to the first or second peak in the experiment
\cite{Baldin}. Proton-proton dibaryon masses are taken from
\cite{Troyan1, Troyan2}.}
\begin{center}
\begin{tabular}
{l@{\hspace{3mm}} c@{\hspace{5mm}} c c} \hline \hline
Reaction  & KAM  & pp-dibaryon masses \cite{Troyan1, Troyan2} \\
\hline
X$+$D$\to$D+D & 1913 & 1916$\pm$2  \\
D$+$X$\to$D+D & 1884 & 1886$\pm$1  \\
D$+$X$\to$X+D & 1886 & 1886$\pm$1  \\
X$+$X$\to$X+D & 1884 & 1886$\pm$1  \\
X$+$X$\to$Y+D & 1886$\to$1898 & 1886$\pm$1, 1898$\pm$1   \\
X$+$D$\to$Y+D & 1916$\to$1884 & 1916$\pm$2, 1886$\pm$1   \\
              & 1965$\to$1937 & 1965$\pm$2, 1937$\pm$2   \\
              & 1980$\to$1953 & 1980$\pm$2, 1955$\pm$2   \\
              & 2106$\to$2086 & 2106$\pm$2, 2087$\pm$3   \\
\hline
D$+$D$\to$X+D & 1965 & 1965$\pm$2   \\
X$+$D$\to$Y+D & 1886$\to$1966 & 1886$\pm$1, 1965$\pm$2   \\
              & 1898$\to$1979 & 1898$\pm$1, 1980$\pm$2   \\
              & 1916$\to$1998 & 1916$\pm$2, 1999$\pm$2   \\
              & 1937$\to$2020 & 1937$\pm$2, 2017$\pm$3   \\
              & 1999$\to$2086 & 1999$\pm$2, 2087$\pm$3   \\
              & 2017$\to$2105 & 2017$\pm$3, 2106$\pm$3   \\
\hline \hline
\end{tabular}
\end{center}
\end{table}
The reactions above the horizontal line explain the first peak and
the reactions below it  explain the second one. It is possible to
verify that the reactions considered for explanation of the data
\cite{Baldin} reproduce masses of all dibaryons observed in refs.
\cite{Troyan1, Troyan2}, with the exception of two of them at
2008$\pm$3 and 2046$\pm$3 MeV/c$^2$.

\section{\label{sec4} An equidistant spectrum assumption}
With an assumption that some of  dibaryons were unrecognized in the
experiments \cite{Troyan1, Troyan2}, it is possible to approximate
the  pp-dibaryon mass spectrum within rather small, at 1 -- 2
MeV/c$^2$ level, experimental errors by the formula
\begin{equation}\label{spectr}
M_n=M_{NN}+10.08\; n,
\end{equation}
where $n=0, 1, 2, ..., 40$, all values are taken in MeV, $M_{NN}$ is
equal to the value of mass of two protons. A quality of this
assumption  is seen, e.g., from a fact that only 4 dibaryons might
be unrecognized in \cite{Troyan1, Troyan2} among the first 14 ones
predicted by (\ref{spectr}).

To check the suggestion of the similarity of pp- and np-dibaryon
mass spectrum, which follows from TABLE~\ref{tab1},  we accepted the
relation (\ref{spectr}) for np-dibaryons too, only changing $M_{NN}$
with the deuteron value of mass. In Tables \ref{tab2} and
\ref{tab3}, the second column specifies masses of ingoing or
outgoing particles, which are allowed by kinematics,
\[
M_Y^2  = M_X^2  + t + M_X P_1 \frac{{\sqrt {t( - 4M_d^2  + t)}
}}{{M_d^2 }}\cos \theta  + \frac{{M_X E_1 t}}{{M_d^2 }},
\]
of the X$+$D$\to$Y+D reaction. Dibaryon masses for the np-system
computed according to (\ref{spectr}) are shown in the third column.

One can see that  each of dibaryons predicted by (\ref{spectr}) in
the range from 1886 to 2198 may contribute to the first or second
peaks, observed in ref. \cite{Baldin}. Thus, new dibaryons predicted
by the equidistant spectrum (\ref{spectr}), taken as an assumption
on basis of \cite{Troyan1, Troyan2},  are also confirmed by the data
\cite{Baldin}. Moreover, quality of the description definitely
improves, since no dibaryon mass calculated using (\ref{spectr}) is
now lost in the description of the data from \cite{Baldin}.

\begin{table} \small
\caption{\label{tab2}\small Kinematically admissible masses (KAM)
which might contribute to the first peak in X$+$D$\to$Y+D reaction.
Dibaryon masses are taken according to the equidistant spectrum
assumption.}
\begin{center}
\begin{tabular}
{l@{\hspace{3mm}} c@{\hspace{5mm}} c c} \hline \hline
Reaction  & KAM  & dibaryon masses, (\ref{spectr})\\
\hline
X$+$D$\to$Y+D & 1916$\to$1884 & 1916, 1886   \\
              & 1926$\to$1895 & 1926, 1896   \\
              & 1936$\to$1905 & 1936, 1906   \\
              & 1946$\to$1916 & 1946, 1916   \\
              & 1956$\to$1927 & 1956, 1926   \\
              & 1966$\to$1938 & 1966, 1936   \\
              & 1976$\to$1948 & 1976, 1946   \\
              & 1986$\to$1959 & 1986, 1956   \\
              & 2047$\to$2024 & 2047, 2027   \\
              & 2057$\to$2034 & 2057, 2037   \\
              & 2067$\to$2045 & 2067, 2047   \\
              & 2077$\to$2056 & 2077, 2057   \\
              & 2087$\to$2066 & 2087, 2067   \\
              & 2097$\to$2078 & 2097, 2077   \\
              & 2107$\to$2087 & 2107, 2087   \\
              & 2118$\to$2099 & 2118, 2097   \\
              & 2128$\to$2109 & 2128, 2107   \\
              & 2138$\to$2120 & 2138, 2118   \\
              & 2148$\to$2131 & 2148, 2128   \\
              & 2158$\to$2141 & 2158, 2138   \\
\hline \hline
\end{tabular}
\end{center}
\end{table}
\begin{table} \small
\caption{\label{tab3}\small Kinematically admissible masses (KAM),
which might contribute to the second peak in X$+$D$\to$Y+D reaction.
Dibaryon masses are taken according to the equidistant spectrum
assumption.}
\begin{center}
\begin{tabular}
{l@{\hspace{3mm}} c@{\hspace{5mm}} c c} \hline \hline
Reaction  & KAM  & dibaryon masses, (\ref{spectr}) \\
\hline
X$+$D$\to$Y+D & 1886$\to$1966 & 1886, 1966 \\
              & 1896$\to$1977 & 1896, 1976 \\
              & 1916$\to$1998 & 1916, 1997 \\
              & 1926$\to$2009 & 1926, 2007 \\
              & 1936$\to$2019 & 1936, 2017 \\
              & 1946$\to$2030 & 1946, 2027 \\
              & 1997$\to$2084 & 1997, 2087 \\
              & 2007$\to$2095 & 2007, 2097 \\
              & 2017$\to$2105 & 2017, 2107 \\
              & 2027$\to$2116 & 2027, 2118 \\
              & 2037$\to$2127 & 2037, 2128 \\
              & 2047$\to$2137 & 2047, 2138 \\
              & 2057$\to$2148 & 2057, 2148 \\
              & 2067$\to$2158 & 2067, 2158 \\
              & 2077$\to$2169 & 2077, 2168 \\
              & 2087$\to$2179 & 2087, 2178 \\
              & 2097$\to$2190 & 2097, 2188 \\
              & 2107$\to$2200 & 2107, 2198 \\
\hline \hline
\end{tabular}
\end{center}
\end{table}

\section{\label{sec5} The dynamical Casimir effect}
The equidistant spectrum regularity observed in \cite{Baldin,
Troyan1, Troyan2} hardly can be  interpreted in the frame of the 6-q
bag model which predicts a different form of spectrum. One may try
to assign it to some kind of oscillator consisting of quarks coupled
by gluon strings \cite{Wang}. However, consideration of the
oscillator wave function with the   constituent quark mass value
indicates that the oscillator should have enormous dimensions. For
example, the state $\psi_{20} (x)$, lying in the middle of the
spectrum observed in\cite{Troyan1, Troyan2}, has the length of about
50 fm.

Actually, it was difficult to find an  explanation better than to
associate the spectrum with the production of pion pairs, strongly
bound to compressed nucleon matter by a deep potential $-U_0$. The
parity conservation requires pions to be produced in pairs (see
below). Therefore, a value of energy of a single pion
\begin{equation}\label{E}
E = \sqrt {p^2  + m^2  - U_0 }
\end{equation}
should be equal to 5.04 MeV $ \equiv  {\rm E}_\pi$.

A meson field  in a rectangular potential well, $ \varphi (\vec r,t)
= e^{ - iEt} \varphi _E (\vec r) $, is described by the Klein -
Gordon - Fock (KGF) steady-state equation,
\[
\frac{1}{{r^2 }}\frac{d}{{dr}}r^2 \frac{{d\varphi _E (r)}}{{dr}} +
(E^2  - m^2  + U_0))\varphi _E (r) = 0
\]
which has a solution $ \varphi _E (r)= A\sin pr/r$ inside the well,
and $ \varphi _E (r)= Be^{ - qr}/r$, $q = \sqrt {m^2 - E^2 }$
outside it. The requirement of continuity of the logarithmic
derivative at the edge of the well, $r=a$, leads to a transcendental
equation
\begin{equation}\label{ctg}
p\;{\rm{ ctg }}(pa) = \sqrt {m^2  - E^2 }
\end{equation}
which is suitable for an estimation of relevant physical values in
the interaction region. Spatial dimensions, corresponding to a given
value of momentum transfer, is \cite{Halzen}
\[
a = \left\langle {r^2 } \right\rangle ^{1/2}   \approx \sqrt 6/
{\rm{ }}\left| {\vec q} \right|  = 0.68 \;{\rm{ fm}}, \qquad \left|
{\vec q} \right|^2   = - t.
\]
Solving eq.~(\ref{ctg}) with this value of $a$, one obtains $ p
 \approx {\rm{0}}{\rm{.53 \; GeV}}, $ and using (\ref{E}), one finds $
\sqrt {U_0 }   \approx 0.55{\rm{ \; GeV}}. $

Touching dynamics of the bound pion production, we suggest that it
is induced by a change of a position of walls forming the potential
well, in  close analogy with emission of electromagnetic waves due
to a motion of resonator's walls. This movement is capable to give
energy to the virtual pions surrounding nucleons and turn them into
real particles, the bound pions. Such a mechanism is known as  the
dynamical Casimir effect, firstly described in \cite{Fulling}. It is
closely connected with the Hawking radiation phenomenon and the
Fulling-Unruh effect \cite{Birrell}. The appeal of this model is it
predicts the meson field with the vacuum quantum numbers, since the
mesons are produced from the vacuum state due to the strong
interaction, conserving all of them. Because of this, the pion field
may be present at the ground state of deuteron, as it follows from
the experimental data\cite{Baldin}, without breaking the deuteron
quantum numbers. As far as the vacuum state has positive parity and
the intrinsic parity of pion is negative, only even number of pions
may be created in the process. Similarly, isospin conservation leads
to a conclusion that pions  may be produced in pairs with $I=0$,
i.e. in the following vector of state:
$$
\Psi _{2\pi }  = \frac{1}{{\sqrt 3 }}(\pi _a^ +  \pi _b^ -   + \pi
_a^ -  \pi _b^ +   - \pi _a^0 \pi _b^0 ).
$$

A picture of the pion production may be  depicted as follows. At
some instant $t_1$ a potential well capable to hold a bound pion
energy level of a value $\varepsilon $ is formed. Then, rather
quickly, the energy level ${\rm E}_\pi > \varepsilon $ is developed
due to a shrinkage of the potential well  in the nucleon collision
process. After that at moment $t_2$, when nucleons is moving away,
the energy level returns to the value $\varepsilon $, and afterwards
it changes again to the Yukawa vacuum, corresponding  $E=0$ and
$q=m$. From mathematical viewpoint, creation of bound pions in this
framework is totally equivalent to the parametric excitation of the
quantum oscillator which appears after the quantization of the
field.

\section{\label{sec6} Pion Bose-Einstein condensate}
The time dependent KGF equation,
\begin{equation}\label{KGF2}
\left[ {\frac{{\partial ^2 }}{{\partial t^2 }} - \frac{{\partial ^2
}}{{\partial r^2 }} + m^2  - U_0 } \right]\psi (r,t) = 0,
\end{equation}
with the evolving boundary conditions gives the wave function inside
the well,
\[
\varphi (r,t) = \chi (t) \sin pr  /r,
\]
where  $\chi (t)$ describes an increasing amplitude of the field
which manifests itself in the  pion production. It  obeys the
equation
\begin{equation}\label{osc}
\frac{{\partial ^2 \chi (t)}}{{\partial t^2 }} + (p^2  + m^2  -
U_0)\chi (t) = 0
\end{equation}
which has the same form as one for a classical oscillator with the
varying frequency $\omega (t) =E(t)$. Therefore, it is possible to
introduce the oscillator Hamiltonian
\begin{equation}\label{Ham}
H = \frac{1}{2}\left( {\pi _\omega ^2  + \omega ^2 (t)\chi _\omega
^2 } \right) = \omega (t)\left( {a_\omega ^ +  (t)a_\omega  (t) +
\frac{1}{2}} \right),
\end{equation}
and draw eq. (\ref{osc}) in the Hamiltonian formalism  framework:
\[
\frac{{\partial H}}{{\partial \pi _\omega  }} = \dot \chi _\omega ,
\qquad - \frac{{\partial H}}{{\partial \chi _\omega  }} = \dot \pi
_\omega,
\]
where
\[
 \chi _\omega   = \frac{{a_\omega   + a_\omega ^ +  }}{{\sqrt
{2\omega } }}, \qquad \pi _\omega   = \frac{{a_\omega   - a_\omega ^
+  }}{{\sqrt {2\omega } }}.
\]

The quantization  may be performed by analogy with the similar
procedure for a quantum field in the box via replacing functions
$a_\omega (t)$ and $a_\omega ^ +  (t)$ by the corresponding
operators. The only non-essential difference is that now the field
does not vanish  at the boundary, but terminates in an exponentially
decaying tail outside the potential well. Fields of this type are
met in solid-state physics \cite{Shockley}. Thus, the quantized
field in the Heisenberg picture is written as
$$
\hat \varphi(r,t) = \hat \chi _\omega  (t) \sin pr/r =  \left(
{\frac{{\hat a_\omega ^ {\dag}  (t) + \hat a_\omega  (t)}}{{\sqrt
{2\omega _1 } }}} \right)\sin pr/r,
$$
for any $t$ in the range of the pion production, $t_1  \le t \le t_2
$. Here $\omega _1 =\omega(t_1)=\varepsilon$. The time evolution of
the field may be expressed in an equivalent form, using Bogoliubov's
canonical transformation (BCT):
\begin{equation}\label{Bog}
\left( {\begin{array}{*{20}c}
   {\hat a(\Delta t)}  \\
   {\hat a^ +  (\Delta t)}  \\
\end{array}} \right) = \overbrace {\left( {\begin{array}{*{20}c}
   {u(\Delta t)} & {v(\Delta t)}  \\
   {u^* (\Delta t)} & {v^* (\Delta t)}  \\
\end{array}} \right)}^{S(\Delta t)}\left( {\begin{array}{*{20}c}
   {\hat a_S}  \\
   {\hat a_S^ + }  \\
\end{array}} \right),
\end{equation}
where $\hat a_S$, $\hat a_S ^ +$ are the annihilation and production
operators in the Schr\"odinger representation, $u(\Delta t)$ and
$v(\Delta t)$ are usual (non-operator) functions. It is obvious that
matrices $S(\Delta t)$  generate a group under multiplication,
$$S( \Delta t)\equiv S(\Delta t_1  +
... + \Delta t_n ) = S(\Delta t_n )...S(\Delta t_1 ).$$ The
commutation relation requirement $[\hat a(t ),\hat a^ + (t )]=1$
leads to a constraint
\begin{equation}\label{su}
\left| u (t)\right|^2 - \left| v (t) \right|^2  = 1
\end{equation}
which means that the group of dynamical symmetry is $ SU(1,1). $

Now we turn to the Schr\"odinger picture and define the group action
in the space of state vectors, rather than in a space of the
parameters describing evolution of operators. Lie algebra of
$SU(1,1)$ is defined by the commutation relations
\[
\left[ {\hat K_1 , \hat K_2 } \right] =  - i \hat K_0 , \; \; \left[
{\hat K_2 , \hat K_0 } \right] = i\hat K_1 , \; \; \left[ {\hat K_0
, \hat K_1 } \right] = i \hat K_2,
\]
or, after introducing
\[
\hat K_ \pm   =  \pm i(\hat K_1  \pm i \hat K_2 ),
\]
by
\[
\left[ {\hat K_0 , \hat K_ \pm  } \right] =  \pm\hat K_ \pm  ,
\qquad \left[ {\hat K_ - , \hat K_ + } \right] = 2\hat K_0.
\]

One can express elements of the $SU(1,1)$ group  through its
generators:
\[
\hat S(dt) = e^{(\beta \hat K_ +   - \beta ^* \hat K_ -   - i\gamma
\hat K_0 )dt}.
\]
But in the case of the Hamiltonian evolution
$$
\hat S(dt)=e^{ - i\hat Hdt},
$$
so that it is possible to rewrite Hamiltonian (\ref{Ham}) in the
form
\[
\hat H = i(\beta \hat K_ +   -  \beta ^* \hat K_ -   - i\gamma \hat
K_0 ).
\]
Corresponding expressions for $\hat K_ +$, $\hat K_ -$ and $\hat
K_0$ are
\[
\hat K_ +   = \frac{{(\hat a^ {\dag}  )^2 }}{2},\; \; \hat K_ -   =
\frac{{\hat a^2 }}{2},\; \; \hat K_0 = \frac{{\hat a\hat a^ {\dag} +
\hat a^{\dag} \hat a}}{4}
\]
for $\pi ^0 \pi ^0$ and
\[
\hat K_ +   = \hat a_ + ^ {\dag} \hat a_ - ^ {\dag} ,\; \hat K_ - =
\hat a_ + \hat a_{ - ,} \; \hat K_0 = \frac{1}{2}(\hat a_ + ^ {\dag}
\hat a_ + + \hat a_ - ^ {\dag} \hat a_ - + 1)
\]
for $\pi ^+ \pi ^-$. In fact, the operators $\hat K_0$ do not lead
to a change of a particle number and it is possible to omit them, at
least for particle number distribution calculations. Thus, the
evolution operator may be defined as an element of the $SU(1,1)$
group of a kind $\hat S(t) = \exp{(\xi \hat K_ + - \xi ^* \hat K_ -
)}.$ Therefore, the state of system at moment $t$ is estimated as
\begin{equation}\label{BCond}
\left| {\psi _t } \right\rangle  = \exp{(\xi \hat K_ + -  \xi ^*
\hat K_ - )} \left| 0 \right\rangle .
\end{equation}

It is possible to notice a similarity of this state to the Glauber
coherent state \cite{Glaub}
$$
\left| {\psi _G } \right\rangle  = e^{\alpha a^ {\dag}   -  \alpha
^* a} \left| 0 \right\rangle = e^{ - \left| \alpha \right|^2 /2}
\sum\limits_{n = 0}^\infty  {\frac{{\alpha ^n }}{{\sqrt {n!} }}}
\left| n \right\rangle
$$
which leads to the Poisson distribution for the probability to find
$n$ particles in the  $\left| {\psi _G } \right\rangle$ state,
$$
w_n  = \left| {\left\langle {n}
 \mathrel{\left | {\vphantom {n {\psi _G } }}
 \right. \kern-\nulldelimiterspace}
 {{\psi _G } } \right\rangle } \right|^2  = e^{ - \left| \alpha  \right|^2 } \frac{{\left|
 \alpha  \right|^{2n} }}{{n!}},{\rm{   }}\left\langle n \right\rangle  = \left| \alpha
 \right|^2 .
$$
Similarly, the state $ \left| {\psi _t } \right\rangle $
reads\cite{Perel}
$$
\left| {\psi _t } \right\rangle  = (1 - \left| \eta \right|^2 )^k
\sum\limits_{m = 0}^\infty  {\left( {\frac{{\Gamma (m +
2k)}}{{m!\Gamma (2k)}}} \right)^{1/2} } \eta ^m \left| {k,k + m}
\right\rangle .
$$
Here  $k$ describes a representations of $SU(1,1)$, $k=1/4$ for
$\pi^0 \pi^0$  and $k = \frac{1}{2}$ for $\pi ^ + \pi ^ -$, $m$ is a
number of pion pairs created, $\eta = \sqrt \rho e^{i\varphi }.$ A
value of $\rho$ may be expressed through the coefficients $u(t_2)$
and $v(t_2)$ of BCT at the end of the pion production, $ \rho =
\left| v \right|^2 / \left| u \right|^2 , $ and $e^{i\varphi }$ is a
phase factor, unessential here. The probability to find $n=2m$
particles in the state is equal to
\begin{equation}\label{wn0}
w_n  = \left| {\left\langle {n}
 \mathrel{\left | {\vphantom {n {\psi _t }}}
 \right. \kern-\nulldelimiterspace}
 {{\psi _t }} \right\rangle } \right|^2  =
 \sqrt {1 - \rho } \frac{{n!}}{{2^n \left[ {\left( {n/2} \right)!} \right]^2 }}
 \rho ^{n/2} ,
\end{equation}
for $\pi ^0 \pi ^0$ system. For $\pi ^+ \pi ^-$, it is
\begin{equation}\label{wnpm}
w_n  = \left| {\left\langle {n}
 \mathrel{\left | {\vphantom {n {\psi _t }}}
 \right. \kern-\nulldelimiterspace}
 {{\psi _t }} \right\rangle } \right|^2 = (1 - \rho )\rho ^{n/2}.
\end{equation}

\section{\label{sec7} Calculation of $\rho$}
The model under consideration  allows to find an exact solution. To
arrive at it, one should only calculate a value of $\rho$. This can
be done in the framework of a certain scattering problem  for a
quantum mechanical particle\cite{Dyk,Popov}, if we accept the usual
scattering matrix formalism assumption: $t_1 \to  - \infty$ and $t_2
\to + \infty$.

In order to make sure of that, let us come back to the Bogoliubov
transformation (\ref{Bog}). One can  see that the coefficients
$u(t)$ and $v(t)$ should satisfy eq. (\ref{osc}), because  the field
should satisfy eq. (\ref{KGF2}), taken in the operator form.
Boundary conditions for the appropriate solutions of (\ref{osc})
follow from requirements
$$
\;\;\hat a (t )\xrightarrow[{t \to  - \infty}]{}\exp{(i \omega_1 t
)}\hat a_S, \qquad \qquad \qquad \qquad
$$
$$
\qquad \hat a (t )\xrightarrow[{t \to  + \infty}]{} C_1 \exp{(i
\omega_1 t)} \hat a_S  + C_2 \exp{(i \omega_1 t)} \hat a_S^{\dag}.
$$
Here the annihilation operator for the outgoing field is taken in
the most general form consistent with its $\exp{(i \omega_1 t )}$
time dependence and the ingoing field operator describes the state
without pions. This implies
$$
\;\; \; u(t)\xrightarrow[{t \to  - \infty}]{} \exp{(i \omega_1 t )},
\;\;\;\;\;v(t)\xrightarrow[{t \to  - \infty}]{}0, \qquad \qquad
\qquad
$$
$$
u(t)\xrightarrow[{t \to  + \infty}]{} C_1 \exp{(i \omega_1 t )}, \;
v(t)\xrightarrow[{t \to  + \infty}]{} C_2 \exp{(i \omega_1 t)}.
$$

Thus, the unknown parameter $\rho$ may be written as
$$\rho (t_2) = \frac{|v(t_2)|^2}{|u(t_2)|^2}=\frac{|C_2|^2}{|C_1|^2}. $$
The requirement (\ref{su}) means that $|C_1|^2$ and $|C_2|^2$ are
not independent. This gives
$$
|C_1|^2 =\frac{1}{1-\rho}, \qquad |C_2|^2 =\frac{\rho}{1-\rho}.
$$
A variable
\[
w(t) = (u(t) +  v(t) ^*)/C_1
\]
also satisfies (\ref{osc}) together with boundary conditions
\begin{equation}\label{wsol}
w(t )\xrightarrow[{t \to  - \infty}]{} e^{  i\omega _1 t } /C_1 ,\;
w(t )\xrightarrow[{t \to  + \infty}]{} e^{ i\omega _1 t }  +
\frac{{C_2^* }}{{C_1 }}e^{-i\omega _1 t }.
\end{equation}

There is a close analogy between eq. (\ref{osc}) for $w(t)$, and its
solution (\ref{wsol}), and the Schr\"odinger equation
$$
\frac{{\partial ^2 \psi (x)}}{{\partial x^2 }} + \left( {\frac{{k^2
}}{{2m}} - V(x)} \right)\psi (x) = 0,
$$
corresponding to the scattering problem  of a particle by a
potential $V(x)$, which has a solution\cite {Landau}
$$
{\rm{e}}^{{\rm{ik}}_{\rm{1}} x}  + B{\rm{e}}^{{\rm{ - ik}}_{\rm{1}}
x}
$$
in the region containing the incident and the scattered wave. In
this framework, the value of $\rho$ corresponds to the reflection
coefficient, $ \rho=R$, of the scattering problem. To achieve the
total mathematical equivalence of the both models, it is necessary
to replace $2m$ by 1 in the Schr\"odinger equation, to transpose
ingoing and outgoing states, and to map:
\[ t \leftrightarrow x, \qquad
\begin{array}{l}
 E^2 (t)  -  V(t) \leftrightarrow k^2 (x)  -  V(  x), \\
 \end{array}
\]
where a time-dependent potential $V(t)$ simulates the changing
boundary conditions. In a simple case when
\[
E(t) = \left\{ {\begin{array}{*{20}c}
   {{\rm E}_\pi = 5.04\;{\rm{ MeV}}{\rm{,  \;\;\;\; for \; 0  <  {\it t}  < }}\tau {\rm{,  }}}  \\
   {\varepsilon {\rm{,  \qquad \;\; for \; 0  >  {\it t}}}{\rm{, \; or \; {\it t}  >  }}\tau } , \\
\end{array}} \right.
\]
one has the scattering by a rectangular potential well of a depth
\[
V_0  = {\rm E}^2 _\pi  - \varepsilon ^2 .
\]
Subject to this proviso, it is possible to find:
\[
\rho  = \frac{1}{{1 + \delta ^2 }}, \qquad{\rm{    }}\delta {\rm{ =
}}\frac{{{\rm{2}}\varepsilon {\rm E}_\pi}}{{V_0 \sin {\rm E}_\pi
\tau }},{\rm{ }}
\]
where $\tau  \sim 1/\Gamma,$ $\Gamma $ is the dibaryon width,
$\varepsilon$ is the only unknown parameter which can be found in
further experiments.  The  data accuracy in \cite{Troyan1, Troyan2}
does not permit to estimate $\varepsilon$ but it allows to conclude
that  $\rho$ is very close to 1, see (\ref{wnpm}) for the registered
value of $n=80$. The distribution (\ref{wn0}) rapidly decreases with
$n$ therefore only the bound  $\pi ^+ \pi ^-$ pairs contribute to
the heavy dibaryon tail observed in \cite{Troyan1, Troyan2}.

\vspace{1cm}
\section{\label{sec8} Discussion and Conclusions}

In the present paper, we confine ourself to consideration of some
experimental evidences for MB production with B=2, leaving aside a
possibility of observation of tribaryons, tetrabaryons,
pentabaryons, etc. One may wonder why so few, if any, signs of
dibaryons exist currently. And particularly, why the partial-wave
analysis (PWA) of N-N elastic scattering did not reveal them. There
are at least two reasonable responses to the second puzzle. First of
all, data  reported by WASA-at-COSY Collaboration \cite{WASA} if
they really inform about the dibaryon natural occurrence mean that a
precision of PWA remains unsatisfactory yet. The second explanation
might be based on a suggestion that some dibaryons in  intermediate
states of the elastic N-N scattering may appear near their mass
shell  only if they are escorted by pions. Corresponding
intermediate states provide  therefore the elastic scattering
amplitude NN $\to$ dibaryon$+$n$\pi$  $\to$ NN with a cut instead a
pole which is usually looked for in PWA. Our suggestion may be
grounded in part by the following reasoning. All dibaryons reported
in \cite{Troyan1, Troyan2, Troyan3} were observed in inelastic N-N
interactions with additional secondary pions. The elastic N-N
scattering amplitude is connected with the inelastic N-N
interactions by the unitarity condition which provides it with all
possible intermediate states. The extra pions  take away an excess
of excitation energy  -- a process which is a some kind of
annealing. This may reconcile two opposite requirements imposed
simultaneously on the system: it must be strongly compressed to form
a compound state and it must be cold enough, since highly excited
levels are usually short-living and elusive.

The second natural question concerns calculations of NN-interactions
below the one-pion threshold in the Chiral Perturbation Theory
(ChPT) framework. Why were there no dibaryons? The  dibaryon with
$M=$ 2.37 GeV stand above one-pion threshold and therefore off this
discussion. As regards light dibaryons, it follows from (5) that a
necessary condition for their existence is $m_{\pi} > 0$. At first
sight, this possibility may be considered in ChPT with the explicit
symmetry breaking. Nevertheless, it is impossible. As it is argued
above, the light dibaryons are an experimental evidence for the pion
Bose-Einstein condensate appearance. It is a purely nonperturbative
effect described by Bogoliubov's transformation which produces a
pion state beyond the range of the Fock space. Perhaps one can find
some traces of this state in ChPT known there as contact terms.
Sometimes they are interpreted  as an evidence for the existence of
the NN-dibaryon vertex, see, e.g., \cite{Ando}. These terms are
introduced if one should describe short-range interactions where a
value of parameter $Q/\Lambda_{\chi}$ is large and the ChPT series
is badly convergent. J.~Soto and J.~Tarr\'us used the same method
for a low energy effective field approximation of QCD for an
explanation of the nucleon-nucleon scattering amplitudes and
obtained an excellent descriptions of the phase shifts \cite{Soto}.

All lattice QCD collaborations  have found stable NN-dibaryons and
dibaryons containing s-quarks, but quark masses in their
calculations are higher than  the physical values, see,
e.g.,\cite{HAL, NPLQCD}. Chiral extrapolations of these results to
the physical point gave, however, evidences against the existence of
such dibaryons, see, e.g., \cite{Shana}. These calculations deal
with ground states and say nothing about unstable states
corresponding to a possibility of two-baryon fusion into 6-quark bag
with a value of mass larger than a sum of masses of the initial
baryons. Recent progress in excited baryon spectroscopy is depicted
in \cite{Lin, Edw}. Corresponding results based on nonphysical quark
masses too cover only one-baryon states so far and are in a poor
agreement with experimental N and $\Delta$ excitation spectra. The
first excited state in two-nucleon system was found in lattice QCD
in \cite{PACS} but with a heavy quark mass corresponding to
$m_{\pi}=0.8$ GeV. Therefore, predicting quasi-bound states of a
multibaryon systems remains a difficult challenge in lattice QCD
till now.

In a paper B.M.Abramov et al \cite{Abramov}, an opinion that
Troyan's resonances were only fluctuations of background was
expressed. In practice, substraction of a background requires a
design of special models, and Yu.A. Troyan elaborated one described
in \cite{Troyan1, Troyan2}. We do not know any explicit objections
against his method, while the solid line in the main figure of the
paper \cite{Abramov} is only an optimal approximation of the
experimental invariant mass spectrum containing, in the general
case, a sum of background and dibaryon contributions. Therefore,
this line cannot be interpreted as the background. It could not be
considered as well as a proof of dibaryon absence by reason of its
smoothness, since usage of more delicate approximations of the
experimental data would reveal a presence of peaks in the spectrum.
Moreover, it is impossible to interpret as statistical fluctuations
peaks shown in Fig.~1 in the paper of Yu.A.~Troyan. Indeed,
statistical fluctuations in one cell of a histogram are Poisson
ones. Therefore, their standard deviation should be equal to
$\sqrt{N}$, where $N$ is a number of events per a cell, shown in
Y-axis in the figure.  It is readily checkable that the fluctuations
near the peak of the histogram overtop substantially the suggested
value. More accurate study of fluctuations with taking into account
experimental errors were performed by Yu.A.~Troyan in
\cite{Troyan2}. He showed that average error  of $M_{pp}$ not far
from the beginning of the spectrum is about 2.4 MeV. This is quite
enough for recognition of isolated dibaryons which are separated
from each other by a distance of 10 Mev. However,  mean correlation
distance $L_c=\Gamma /2$, of the fluctuations identified as
dibaryons at small values of $M_{pp}$ is of the same order. This
implies that the true resonant widths of the dibaryons should might
be less than those seen in Fig.~1 in \cite{Troyan2} and, actually,
the peaks might be higher than they appear in the figure. Therefore,
very small probabilities of the dibaryons might be a maverick, found
in \cite{Troyan2}, seem to be rather realistic. To confirm this
suggestion future experiments  must have resolution at least at a
level 1 MeV due to higher statistics and less experimental errors.

There is another reason might explain the difference between
Yu.A.~Troyan and B.M.~Abramov et al experiments. As it was suggested
in our paper, observation of dibaryons is possible only under the
conditions of "deep cooling". Let us compare. Only a reaction $pn
\to pp \pi^-$ was considered in the paper of B.M.Abramov et al.
Reactions investigated by Yu.A.~Troyan include: $pn \to pp \pi^-$,
$pn \to pp \pi^- \pi^0$, $pn \to pp \pi^+ \pi^- \pi^-$, $pn \to pp
\pi^+ \pi^- \pi^- \pi^0$. We can see from kinematics, and explicit
comparison of the data from \cite{Troyan1, Troyan2} and
\cite{Abramov}, that the effective mass spectrum is hotter indeed in
Abramov's experiment. The Bose-Einstein condensate may not arise at
such conditions. Therefore, one might suggest that the first
reaction from the Troyan's list gave only a noise to the dibaryon
signal observed. And we see, indeed, that the tail of distribution
in Fig.1 in the paper of Yu.A.~Troyan \cite{Troyan1, Troyan2}
contains visible strips in which the fluctuations are symmetrical
against the background. This may be a signature of a small dibaryon
contribution in this region.

Our consideration of the data on the hard deuteron-deuteron
scattering \cite{Baldin}   meets the expectation to observe the
transition of nucleon matter into other states using the method of
deep cooling which allows to recognize quasi-resonance peaks in the
reaction cross-section. One of them shown in Fig.~\ref{Peak3} is
very close to a dibaryon reported by WASA-at-COSY Collaboration
\cite{WASA} at 2.37 GeV, $\Gamma\approx$ 70 MeV. Another one, at a
short distance from 2.5 GeV, may be explained by  the sum of
contributions of N$+$D$\to$N$^*+$D reactions, see Fig.~\ref{Peak3}.
As far as this explanation is far from being perfect, it is possible
to suspect also the existence of another dibaryon therein.

As concerns the dibaryons obeying the equidistant spectrum
regularity observed in \cite{Baldin, Troyan1, Troyan2}, they hardly
can be interpreted in the frame of the 6-q bag model. It is very
likely  to assign them to the production of pion pairs strongly
bound to compressed nucleon matter. The analysis of the data from
\cite{Baldin} reveals the possibility of presence of the pion
Bose-Einstein condensate in the ground state of deuteron, see
(\ref{BCond}). According to this analysis, the condensed pion field
in deuteron can change in hard nuclear collisions. The pion
Bose-Einstein condensate might also appear in the compressed
proton-proton system subjected to a proper cooling, according to the
experimental hints from \cite{Troyan1, Troyan2}. The theory predicts
the characteristic mass distribution for dibaryons of this type,
which may be considered as an experimentally feasible signature of
the pion Bose-Einstein condensate.

It is reasonable to ask whether the pion Bose-Einstein condensate
arises in compressed $k$-nucleon systems for $k>2$. If this is true,
it can impact essentially on collective flows at the final stage of
high-energy nuclear collisions, especially on the sideflow
\cite{Herrmann}.

It should be noted that the state of  pion field (\ref{BCond}) has a
mathematical and physical prototype in quantum optics, known there
as the squeezed vacuum \cite{Walls}. Using this interpretation, one
may qualify the operator $\hat S(t) = \exp{(\xi \hat K_ + - \xi ^*
\hat K_ - )}$ defined above as the squeeze operator. An appropriate
squeeze factor $r$  can be expressed through the expectation value
of the pion number in this state: $ \sinh^2 r = \left\langle \hat
a^{\dag} \hat a \right\rangle$ for $\pi^0 \pi^0$ and $ \sinh^2 r =
\left\langle \hat a_{\pm}^{\dag} \hat a_\pm \right\rangle$ for
$\pi^+ \pi^-$ pairs.

\end{document}